\documentclass[twocolumn,showpacs,preprintnumbers,amsmath,amssymb,eqsecnum,pre]{revtex4}

\usepackage{graphicx}      
\usepackage{dcolumn}       
\usepackage[usenames,dvipsnames]{color}

\newcommand{\beq}{\begin{equation}}
\newcommand{\eeq}{\end{equation}}
\newcommand{\beqn}{\begin{eqnarray}}
\newcommand{\eeqn}{\end{eqnarray}}

\newcommand\beqa{\begin{eqnarray}}
\newcommand\eeqa{\end{eqnarray}}

\newcommand{\de}{\delta}

\newcommand{\la}{\langle}
\newcommand{\ra}{\rangle}

\definecolor{verde}{rgb}{.1,.4,0}
\definecolor{violet}{rgb}{.5,.0,1}
\definecolor{gray}{rgb}{.5,.5,.5}
\definecolor{brown}{rgb}{.4,.1,0}
\definecolor{orange}{rgb}{1,.5,0}
\definecolor{bordo}{rgb}{.5,0,.2}
\definecolor{rojo}{rgb}{.8,0,0}
\definecolor{otro}{rgb}{.9,0,0}

\newcommand{\ed}{\end{document}}

\begin{document}
\title{Conditional pair distributions in many-body systems:\\ 
       Exact results for Poisson ensembles}

\author{Ren\'e D. Rohrmann}
\email{rohr@icate-conicet.gob.ar}
\homepage{http://icate-conicet.gob.ar/rohrmann}
\affiliation{Instituto de Ciencias Astron\'omicas, de la Tierra y del
Espacio, UNSJ-CONICET, Av.\ Espa\~na 1512 Sur, 5400 San Juan, Argentina}
\author{Ernesto Zurbriggen}
\email{ernesto@oac.uncor.edu}
\affiliation{Observatorio Astron\'omico, Universidad Nacional de C\'ordoba,
Laprida 854, 5000 C\'ordoba, Argentina}

\date{\today}
\begin{abstract}
We introduce a conditional pair distribution function (CPDF) which 
characterizes the probability density of finding an object (e.g., a particle 
in a fluid) to certain distance of other, with each of these two having a 
nearest neighbor to a fixed but otherwise arbitrary distance. 
This function describes special four-body configurations, but also contains 
contributions due to the so-called mutual nearest neighbor (two-body) and 
shared neighbor (three-body) configurations. The CPDF is introduced to improve 
a Helmholtz free energy method based on space partitions.
We derive exact expressions of the CPDF and various associated quantities
for randomly distributed, non-interacting points at Euclidean spaces of one, 
two and three dimensions. Results may be of interest in many diverse 
scientific fields, from fluid physics to social and biological sciences.

\end{abstract}


\maketitle
\section{ Introduction} \label{s.intr}

Spatial distributions of objects are ubiquous in nature. The pattern of 
distribution of a population can be characterized by statistical relationships 
between each component and its nearby companions. 
Methods based on neighbor relationships and two-point correlations are some of 
the simplest and most popular techniques for statistical pattern recognition.
The analysis of spatial patterns is especially useful in modeling and 
describing a wide variety of natural \cite{CE54,Ro69,Bu43,Te94,SC99,CE55,
BO70,Ge08} and social phenomena \cite{CJ04,MD41,Ba77,BG98,MC99}.

Methods for spatial structure studies of many-body systems have been developed 
and extensively applied in physics of condensed matter \cite{Ba75,HM06}. 
Similar techniques are employed to understand the associations of galaxies and 
the large scale structures of the Universe \cite{Li89,Pe93}.
The main spatial structure function in the study of liquids is the pair
distribution function (PDF). For statistically homogeneous and isotropic 
systems, which are the focus of this paper, the PDF is a radial function 
$g(r)$, traslationally invariant, which characterizes the probability density 
of finding a particle at some given distance $r$ from a reference particle. 
Assuming pairwise particle interactions, $g(r)$ can be used to obtain 
equations involving macroscopic thermodynamic variables \cite{HM06}. 
An additional description can be obtained from $k$th-nearest-neighbor 
PDFs $g_k(r)$ \cite{MB86,Ed91,Ma92,Ke99,Bh03}.
The first nearest neighbor distribution, $g_1(r)$, is particularly useful to 
understand local fields around a particle in a fluid \cite{He09,RC74,Ma81,
TL90}, or a star in a stellar cluster or galaxy \cite{Ch43}, and it has 
applications in other problems in physics \cite{Ma59}.
However, a further detailed microscopic information is sometimes required to 
statistically characterize physical properties of fluids and materials. 
For instance, mutual-nearest-neighbor (MNN) pair descriptors 
are used in the study of solute relaxation processes in fluids \cite{LS98}, 
phonon spectra analysis of disordered systems \cite{Wu05}, and for explaining 
the behavior of glass-forming liquids \cite{PG05}. In general, the probability
that particles in a fluid form MNN pairs can not be exactly deduced from 
the standard $g(r)$ or individual PDFs $g_k(r)$. Therefore, a superior spatial 
structure descriptor is highly desirable.

We propose here the introduction of a more complete description of the 
microscopic neighborhood structure around a reference particle.
We introduce a conditional pair distribution function (CPDF) 
which gives the probability density of finding a particle to a certain 
distance from other in a fluid, with each of these two having a nearest 
neighbor to a fixed but otherwise arbitrary distance. 
This function describes special four-body configurations, but also contains 
contributions due to the mutual nearest neighbor (two-body) and shared 
neighbor (three-body) configurations. 
The new PDF is a spatial descriptor at the two-point distribution level 
but contains nontrivial higher-order structural information. 
In particular, it is a superior descriptor sensitive to topological
connectedness information of the point pattern and could resolve a variety 
of degenerate point configurations associated with the standard PDF $g(r)$
\cite{JS10}. 

The primary purpose of introducing the CPDF is to establish a correct
expression for evaluating the interparticle potential energy in the so-called
thermostatistical space partition (TSP) formalism for fluids. 
The TSP approach, which was introduced in Ref. \cite{Ro05} for the case of 
pure substances and extended to mixtures in Ref. \cite{RZ06}, was motivated by 
difficulties with the calculation and qualitative prediction of optical 
properties in white dwarf stars, currently based on the occupation probability 
formalism of Hummer \& Mihalas \cite{HM88} and whose consistency has been 
discussed in Ref. \cite{Ro02}. 
The TSP method combines the free-energy minimization technique \cite{Ga69}
with space partitions that asign to each particle a volume $v$,
which is determined by the distance of the particle to its closest neighbor,
and roughly measures the free space surrounding each particle.
The CPDF introduced in the present paper, denoted $g_{vv'}(r)$, is a PDF where 
the volumes $v$ and $v'$ of both pair particles are specified. 
It will be shown here that this CPDF is needed for an appropriate 
representation of the potential energy of fluids within the TSP formalism.

Explicit evaluation of the CPDF $g_{vv'}(r)$ is considerably more complicated 
than the computation of the conventional PDF $g(r)$. 
In the pairwise additive assumption for the potential energy of a fluid, 
the CPDF can be evaluated using a hierarchy of integral equations for $s$-body 
correlation functions, for instance the Yvon-Born-Green hierachy \cite{HM06},
truncated by introducing some appropriate closure approximation.
In the general case, with $g_{vv'}(r)$ representing four-particle 
configurations, the hierarchy expansion can be truncated in factors with 
correlation functions higher than $s=5$ (in analogy with the mean-field 
approximation used in Ref. \cite{LS98} to assess the MNN pair distribution 
of Lennard-Jones fluids). A method of calculating $g_{vv'}(r)$ by means of 
this scheme is currently being prepare in detail. 
In the present contribution we will restrict ourselves to the evaluation
of $g_{vv'}(r)$ for random, non-interacting systems or so-called Poisson 
ensembles, which can be directly derived from simple and well-known  
probabilistic properties (e.g., \cite{Fr55,Gi62,Ro69}).
The aim is two-fold. First, the study of the CPDF for random 
points may prove to be a useful guide for investigating this function in
other systems, apart from its own importance at a fundamental level.
Second, Poisson statistics is an especially powerful tool for studies of
a wide variety of systems, and its CPDF may provide a new method of analysis.
In fact, Poisson statistics emerges as a useful alternative in view of the 
practical limitations of a rigorous treatment for the complex natural systems.
Random samples are employed very widely by comparison or contrast in the
analysis of observed event data \cite{Di03}, or as a convenient first 
approximation to describe the spatial distribution of a variety of natural
systems, from biological species \cite{CE54,Ro69,Bu43,Te94,SC99,CE55,BO70,Ge08} 
to stars on the sky \cite{Ba81,Go93,Lo10}.
The present work is part of an effort dedicated to improving the TSP formalism 
and to providing a well defined and very demanding test model of the CPDF for 
further research.

The paper is structured as follows. In Section \ref{s.def} the conditional 
pair distribution function is formulated and general results are presented. 
Its implication within the TSP formalim is described in Section \ref{s.TSP}.
Section \ref{s.pe} contains a derivation of the CPDF for Poisson patterns in 
one, two, and three spatial dimensions. Some derived results are examined 
there. Concluding remarks are contained in Section \ref{s.co}.

\section{Definitions and general relations} \label{s.def}

Consider a system of $N$ identical particles distributed in a $D$-dimensional
Euclidean space with volume $V$. For conciseness, the radial distance $r$ 
from a reference particle will be expressed in terms of the volume 
$\omega=\alpha r^D$ of a $D$-dimensional sphere of radius $r$ centered in 
the particle, with
\beq \label{e.alpha}
\alpha \equiv \frac{\pi^{D/2}}{\Gamma(1+\frac D2)},
\eeq
where $\Gamma$ is the gamma function.
The conventional PDF, henceforth expressed as $g(\omega)$, describes the 
probability density of finding a particle on the surface of a sphere of 
volume $\omega$ centered in the reference particle. 
More exactly, $g(\omega)$ is defined such that
\beqn
 n g(\omega)d\omega\equiv && \text{mean number of particles between the}\cr
 &&\text{surfaces of $\omega$ and $\omega+d\omega$ centered at the}\cr 
 &&\text{reference particle},
\eeqn
where $n=N/V$ is the mean density in the fluid. 

It is assumed here that for any reference particle there is one and only one 
nearest neighbor. This means that if two or more particles are equidistant 
from the reference, one of them will be arbitrarily designated as the first 
neighbor. The distance $a$ between a particle and its nearest 
neighbor will be expressed by the volume $v$ of a $D$-dimensional sphere of 
radius $a$. 
The volume $v$ will be referred to as the {\em available volume} (AV) of the 
particle because it roughly measures the free-particle space surrounding it. 
The size distribution of AV in the fluid is given by the distribution $n_v$, 
where $n_vdv$ is the number density of particles with available volume 
between $v$ and $v+dv$. The quantity $(n_v/n)dv$ is clearly the probability
that the nearest neighbor lies in the volume $dv$ of a spherical shell 
centered in the reference particle. Therefore, $n_v/n$ is equivalent to the 
so-called {\em first neighbor distribution} \cite{RC74,Ma81,TL90}.
In the thermodynamic limit ($N,V\rightarrow \infty$, keeping $N/V$ constant), 
it follow that
\beq \label{e.nnv}
n=\int_0^\infty n_v dv.
\eeq
%

%
\begin{figure}
\scalebox{0.43}{\includegraphics{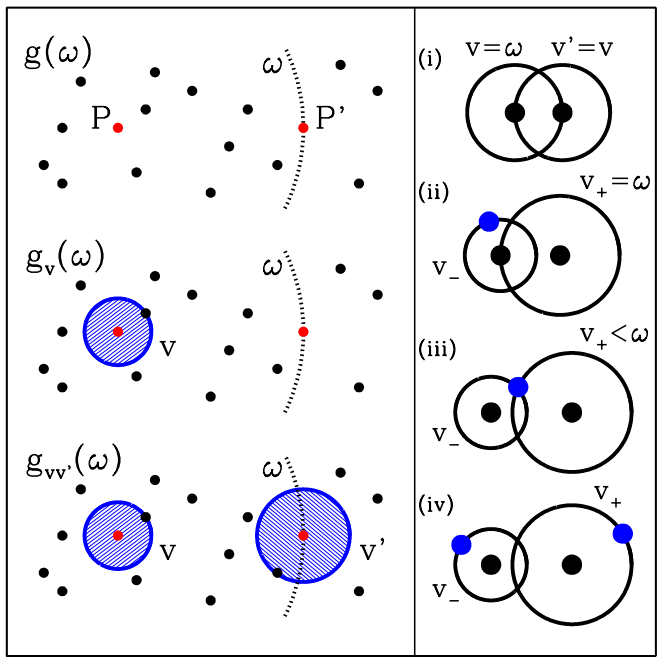}}
\caption{\label{f.fggg} (Color online) 
{\it Left:} Example of a point pattern with a characterization of the 
evaluation of the pair distribution functions $g(\omega)$, $g_v(\omega)$ and 
$g_{vv'}(\omega)$.
The point $P'$ lies in the surface of the spherical volume $\omega$ centered 
at the reference point $P$. Nearest neighbors of $P$ and $P'$ lie in the 
surfaces of spherical volumes $v$ and $v'$ centered in $P$ and $P'$, 
respectively.
{\it Right:} Representation of the four classes of ($v,v'$) pairs (see text).
Here, $v_+\equiv \max(v,v')$ and $v_-\equiv \min(v,v')$.
}
\end{figure}
%
The conditional pair distribution functions $g_v(\omega)$ and $g_{vv'}(\omega)$
are defined by the following quantities
\beqn \label{gv.def}
 n g_v(\omega)d\omega\equiv && \text{mean number of particles between the}\cr
 &&\text{surfaces of $\omega$ and $\omega+d\omega$ centered at the}\cr 
 &&\text{reference particle with AV $v$},
\eeqn
\beqn \label{gvv.def}
 n_{v'} g_{vv'}(\omega)dv'd\omega\equiv 
 &&\text{mean number of particles with  AV}\cr
 &&\text{between $v'$ and $v'+dv'$ lying between} \cr
 &&\text{the spherical volumes $\omega$ and $\omega+d\omega$}\cr
 &&\text{centered at a particle with AV $v$.}\cr
 &&
\eeqn 
According to the meaning of $n_v$, it is clear that $g(\omega)$ is just a 
weighted average of $g_v(\omega)$ over the whole range of $v$, that is
\beq \label{ggv}
n g(\omega) = \int_0^\infty n_{v}g_{v}(\omega)dv.
\eeq
Similarly, the integration of (\ref{gvv.def}) over $v'$ covering the whole 
range of available volume gives (\ref{gv.def}), i.e.,
\beq \label{gvgvv}
n g_v(\omega) = \int_0^\infty n_{v'}g_{vv'}(\omega)dv'.
\eeq
From (\ref{ggv}) and (\ref{gvgvv}) it follows that
\beq \label{ggvv}
n^2 g(\omega) = \int_0^\infty \int_0^\infty n_v n_{v'} g_{vv'}(\omega)dv dv'.
\eeq

From the definitions given above, we can see that the function $g_v(\omega)$ 
is the probability density that a particle lies at the surface 
of a spherical volume $\omega$ centered in a randomly chosen particle 
{\em with} AV $v$. Similarly, $g_{vv'}(\omega)$ is the probability density of 
finding a particle with AV $v'$ in the surface of a spherical volume $\omega$ 
centered in a particle with AV $v$.
The conditional character of the probability densities $g_v(\omega)$ and
$g_{vv'}(\omega)$ appear due to the specification of the AV
$v$ and $v'$ for one or both particles of each pair.
Due to the symmetry of the problem, the function $g_{vv'}(\omega)$ is 
symmetric in the variables $v$ and $v'$. 
Fig. \ref{f.fggg} ({\it left}) displays a schematic point pattern with
characterizations of the standard PDF $g(\omega)$ and the CPDFs $g_v(\omega)$ 
and $g_{vv'}(\omega)$. The function $g_{vv'}(\omega)$ and the distribution
$n_v$ are the central quantities of the present issue since knowledge of them 
allows one to calculate $g_v(\omega)$ and $g(\omega)$ with (\ref{gvgvv}) 
and (\ref{ggvv}).

\begin{table}
\vspace{-0.5cm}
\caption{Classes of pairs ($v$,$v'$) and some properties.} \label{t.1}
\vspace{-0.cm}
\begin{ruledtabular}
\begin{tabular}{ll|l|l}
 & Configuration     &  n-body  & $\omega$-range \\
\hline
(i)   &mutual nearest neighbor &  2 bodies & $\omega=v=v'$\\
(ii)  &simple nearest neighbor &  3 \quad '' & $\omega=v_+\equiv \max(v,v')$\\
(iii) &shared nearest neighbor &  3 \quad '' & $v_+<\omega<\omega_{vv'}$\\
(iv)  &other                   &  4 \quad '' & $v_+<\omega$\\
\end{tabular}
\end{ruledtabular}
\end{table}
%
The function $g_{vv'}(\omega)$ describes in general four-body 
configurations, i.e., pairs ($v$,$v'$) composed of two particles with a 
third particle in the surface of the volume $v$ centered in the first one 
and a fourth particle in the surface of $v'$ centered in the second one. 
However, there are also other configurations of pairs ($v$,$v'$) formed 
by only two or three particles.
As is illustrated in Fig. \ref{f.fggg} ({\it right}) and detailed in 
Table \ref{t.1}, there are four independent classes of pairs ($v,v'$): 
\begin{itemize}
\item{\em Class} (i) denotes mutual nearest neighbors, i.e., each
particle of the pair ($v$,$v'$) is the nearest neighbor of the other.
\item{\em Class} (ii) corresponds to the configuration where one and only one 
particle of the pair ($v$,$v'$) is the nearest neighbor of the other.
\item{\em Class} (iii) refers to the shared nearest neighbor configuration, 
i.e., both particles of the pair ($v$,$v'$) have the same nearest neighbor.
\item{\em Class} (iv) denotes pairs ($v$,$v'$) which involve a configuration
with four different particles.
\end{itemize}
Because classes (i-iv) are mutually exclusive and exhaustive, we have
\beq \label{e.gvv}
g_{vv'}(\omega) = g^{(i)}_{vv'}(\omega)   + g^{(ii)}_{vv'}(\omega) 
                 +g^{(iii)}_{vv'}(\omega) + g^{(iv)}_{vv'}(\omega), 
\eeq
where $g_{vv'}^{(\nu)}(\omega)$ is the contribution to $g_{vv'}(\omega)$ due
to pairs ($v$,$v'$) of class ($\nu$).

Geometrical considerations which involves the available volume $v$ of a 
particle or, equally, the distance to its nearest neighbor, can be employed 
to derive some general properties of the contributions to $g_{vv'}(\omega)$ 
on the right hand side of (\ref{e.gvv}).
As can be seen from Fig. \ref{f.fggg}, pairs class (i) only occurs for 
$\omega=v=v'$ and, therefore, the mathematical representation of the term 
$g_{vv'}^{(i)}(\omega)$ must have singularities. Indeed, taking into account 
(\ref{gvv.def}), it is easily demonstrated that this term is proportional to 
the product of two Dirac delta functions, 
\beq \label{e.gi}
g_{vv'}^{(i)}(\omega) \propto \delta(v-v')\delta(\omega-v).
\eeq
A similar behavior is deduced for the second term on the right-hand side of 
(\ref{e.gvv}). Pairs ($v$,$v'$) of class (ii) occur only for $\omega=v_+$, 
with 
\beq
v_+ \equiv \max(v,v'), 
\eeq
and therefore
\beq \label{e.gii}
g_{vv'}^{(ii)}(\omega) \propto \delta(\omega-v_+).
\eeq
Since two particles cannot share the nearest neighbor when their volumes $v$ 
and $v'$ do not overlap, the third term on the right of (\ref{e.gvv}) 
vanishes for $\omega>\omega_{vv'}$, where 
\beq \label{wvv}
\omega_{vv'} \equiv (v^{1/D}+v'^{1/D})^D.
\eeq
The third term vanishes also at $\omega<v_+$ because no particle lies at 
lower distance than the nearest neighbor. Thus
\beq
g_{vv'}^{(iii)}(\omega) = 0, \quad \omega \notin (v_+,\omega_{vv'}).
\eeq
Similarly, the fourth term vanishes for $\omega<v_+$,
\beq \label{e.giv0}
 g_{vv'}^{(iv)}(\omega) = 0, \quad \omega < v_+.
\eeq
Therefore, $g_{vv'}(\omega)$ vanishes at $\omega<v_+$ and has singularities
for pair configurations with $\omega=v=v'$ and $\omega=v_+$. 

At the fluid context and in analogy to the standard PDF, one may expect 
that, at sufficiently large interparticle distances (large $\omega$), pairs 
($v$,$v'$) are uncorrelated and the CPDF must then reduce to unity 
\beq
\lim_{\omega\rightarrow \infty} g_{vv'}(\omega) = 1 .
\eeq
%

\section{The role played by $g_{vv'}(\omega)$ in the TSP formalism} 
\label{s.TSP}

Let us define $u_{vv'}(\omega)$ as the interparticle potential between two
particles forming a pair ($v,v'$), with a particle centered in the volume 
$\omega$ and another one located on the surface of $\omega$. The potential 
$u_{vv'}(\omega)$, where the AV $v$ and $v'$ for both pair particle are 
specified, is a generalization of the usual pairwise interparticle potential 
$u(\omega)$. $u_{vv'}(\omega)$ formally takes into account 
perturbative effects on the two-body interaction potential introduced by the 
nearest neighbor of each particle in the pair. Many-body contributions to
effective pair potentials are known for liquid hydrogen \cite{RB74, Ro83, He90}
and have been theoretically considered for fluid helium \cite{AC94}. 
However, potential energy surfaces of four-particle systems are needed for 
a full knowledge of $u_{vv'}(\omega)$ in a fluid. For the moment, 
{\it ab initio} potential evaluations are limited to three-body systems at few 
geometrical configurations and for some atomic, molecular and colloidal 
systems \cite{WR96, So03, Ru05}.

Multiplication of (\ref{gvv.def}) with $u_{vv'}(\omega)$ and integration over
all $\omega$, $v$ and $v'$, yields the total potential energy of the system
\beq \label{i.ug2}
\Phi=\frac{1}{2V}\int_0^V\int_0^V N_vN_{v'} 
      \left[\int u_{vv'}(\omega) g_{vv'}(\omega)d\omega\right]  dv' dv,
\eeq
where $N_v=n_v V$, $N_{v'}=n_{v'} V$, and the factor $1/2$ avoids the double 
count of pairs ($v,v'$). If one writes
\beq  \label{uv}
u_v(\omega)=\frac {\int n_{v'}u_{vv'}(\omega)g_{vv'}(\omega)dv'}
{\int n_{v'}g_{vv'}(\omega)dv'},
\eeq
then (\ref{i.ug2}) can alternatively be written as
\beq \label{i.ug1}
\Phi=\frac{N}{2V}\int_0^V N_v \left[\int u_v(\omega) g_v(\omega) 
d\omega\right] dv.
\eeq
Similarly, if we define an averaged pair potential as 
\beq \label{u}
u(\omega)=\frac {\int n_vu_v(\omega)g_v(\omega)dv}
{\int n_vg_v(\omega)dv},
\eeq
then the total potential expression reduces to that typically used in physics 
of liquids, which is based on the conventional PDF $g(\omega)$, i.e.,
\beq \label{i.ug}
\Phi=\frac{N^2}{2V}\int u(\omega) g(\omega) d\omega.
\eeq
For systems with two-particle potential without many-body effects, 
$u_{vv'}(\omega)=u_v(\omega)=u(\omega)$. Otherwise, the CPDF $g_{vv'}(\omega)$
can be used to calculate {\it effective} pair potentials 
[Eqs. (\ref{uv}) and (\ref{u})] directly from {\it ab initio} many-body 
potentials $u_{vv'}(\omega)$ (when these become available), and thus the 
conventional evaluation of the total potential energy in liquid theory is 
preserved [Eq. (\ref{i.ug})].

The thermostatistical space partition (TSP) formalism is a mathematical model 
for moderately dilute gases which combines the free-energy minimization method 
\cite{Ga69} with space partitions. It was developed for one-component 
\cite{Ro05} and multi-component \cite{RZ06} fluids. 
Applications to simple gas models (hard spheres, van der Waals fluids, 
attractive particles, and partially ionized hydrogen gas) have shown that TSP 
provides a useful tool to evaluate self-consistently thermodynamic and 
structural properties of gases at low densities as found in white dwarf 
atmospheres, where conventional techniques fail to reproduce gas opacities 
affected by non-ideal effects (e.g., the Lyman continuum opacity \cite{Be97, 
Ro02}). 
The novel feature of the TSP formalism is that it allows one to evaluate, in a 
thermodynamically self-consistent way, populations of atoms and molecules in 
internal energy states  as a function of the available volume $v$, a variable 
linked directly with different degrees of interparticle perturbation.

For one-component fluids, the TSP formalism is defined in terms of a Helmholtz 
free-energy model $A$ based on the occupation number distribution $N_v$ 
of the variable $v$ \cite{Ro05},
\beq \label{e.A}
A=\int_0^V \frac{N_v}{\beta} \ln\left(\frac{N_v\lambda^D}{N}\right)dv+\Phi,
\eeq
where $\beta=1/(kT)$, $\lambda=\sqrt{h^2\beta/(2\pi m)}$, $D$ is the space
dimension, $k$ the Boltzmann constant, $h$ the Planck constant, $m$ the 
particle mass and $T$ the temperature. The first term on the right hand side 
of Eq. (\ref{e.A}) is the ideal free energy as given by the TSP formalism.
The space distribution $N_v$ for a particle system at equilibrium is 
determined by minimization of $A$ subjected to total particle number 
and total volume  constrains (see \cite{Ro05} for details),
\beq \label{c.N}
N=\int_0^V N_vdv, 
\eeq
\beq \label{c.V}
V=\int_0^V vN_vdv.
\eeq
In Ref. \cite{Ro05}, the interparticle interactions in the fluid were 
incorporated into the TSP approach via the potential $u_v(\omega)$ and the PDF 
$g_v(\omega)$. Therefore, the total potential energy $\Phi$ was written in
\cite{Ro05} as (\ref{i.ug1}), but assuming that $u_v(\omega)$ is
{\it independent} of $N_v$. In such case, $\Phi$ is a {\it linear} functional 
of $N_v$ and the minimization free energy procedure yields
\beq \label{e.Nv1}
N_v =\frac{N}{\lambda^D} \exp\left[-\beta\left(\gamma v+\frac{\phi_v} 2 
+\frac{\Phi}{N}-\mu\right)\right],
\eeq
with
\beq \label{e.phi1}
\phi_v= n\int u_v(\omega)g_v(\omega)d\omega
\eeq
where $\mu$, the chemical potential, and $\gamma=P-\Phi/V$, the pressure 
minus the potential energy density, are the Lagrange multipliers for 
our constrains.

\begin{figure}
\scalebox{0.35}{\includegraphics{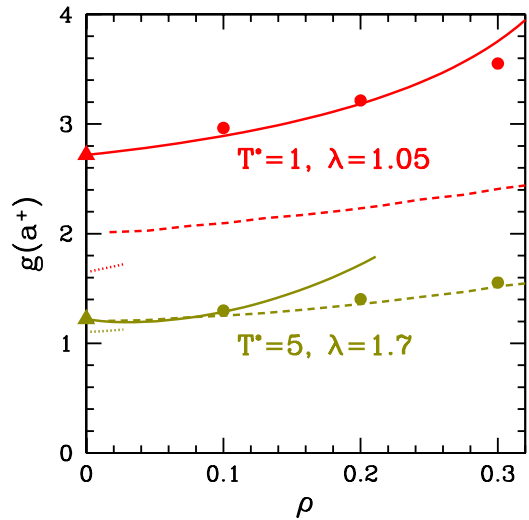}}
\caption{\label{f.adhe} 
(Color online)
Contact values $g(a^+)$ of the PDF for attractive hard spheres in the TSP 
formalism based on Eq. (\ref{e.Nv2}) (solid lines), are compared with exact 
values in the limit of zero density (triangles), Monte Carlo simulations 
\cite{La05} (circles), results of a perturbative theory \cite{TL94} (dashed 
lines), and results derived from Eq. (\ref{e.Nv1}) (dotted lines).
Here $T^*=kT/\epsilon$ is the reduced temperature, $\rho=n d^3$ the reduced
number density, and $\lambda$ the potential range in units of the particle 
diameter $d$ [i.e., $\xi=a(\lambda^3-1)$ and $a=4\pi d^3/3$ in 
Eq. (\ref{a.adh})].
}
\end{figure}

However, if we adopt the more general potential energy expression given by 
(\ref{i.ug2}), $\Phi$ is a {\it quadratic} functional of $N_v$. In this case,
it is easily deduced that the minimization technique leads to
\beq \label{e.Nv2}
N_v =\frac{N}{\lambda^D} \exp \left[-\beta(\gamma v+\phi_v -\mu)\right],
\eeq
with $\phi_v$ and $u_v(\omega)$ respectively given by (\ref{e.phi1}) and 
(\ref{uv}) or, equivalently,
\beq
\phi_v= \int \int n_{v'} u_{vv'}(\omega)g_{vv'}(\omega)dv'd\omega.
\eeq
The main difference between Eqs. (\ref{e.Nv1}) and (\ref{e.Nv2}) is the 
dependence of $N_v$ on $\phi_v$.
It is important to note that the results obtained in \cite{Ro05} from the TSP 
approach for hard systems and van der Waals fluids remain unchanged if one
uses (\ref{e.Nv2}) instead (\ref{e.Nv1}). This is because of the special form
of the two-particle potentials for these systems. However, Eq. (\ref{e.Nv2}) 
is the correct one and must be used in general. In particular, 
Eq. (\ref{e.Nv2}) is supported by results corresponding 
to attractive hard spheres (AHS). The implentation of the TSP approach to the
AHS model is shown in Appendix \ref{a.TSP}. We obtain the following contact 
value of the conventional PDF for this model,
\beq \label{a.gcont}
g(\omega=a^+)=\frac{\beta\gamma}{nB} e^{\beta\epsilon(1+n\xi)},
\eeq
with $\beta\gamma$ and $B$ given by Eqs. (\ref{a.bg}) and (\ref{a.B}).
A density expansion of (\ref{a.gcont}) yields
\beq
g(a^+)=e^{\beta\epsilon}\left\{1+\left[b+(2+\beta\epsilon
-2e^{\beta\epsilon})\xi\right]n+O(n^2)\right\},
\eeq
in agreement with the value expected at zero density 
$g(a^+)=e^{-\beta u(a^+)}=e^{\beta \epsilon}$. On the contrary, it may be
deduced that the contact value at zero density resulting from 
Eq. (\ref{e.Nv1}) is incorrectly $e^{\beta\epsilon/2}$.
Although the TSP approach is being developed mainly for the low densities
found in stellar atmospheres, Fig. \ref{f.adhe} shows that Eq. (\ref{a.gcont}) 
can also reproduce satisfactorily the results at moderately low densities 
obtained from Monte Carlo simulations \cite{La05}.
In conclusion, the correct expression (\ref{e.Nv2}) for the space distribution
$N_v$ at equilibrium in the TSP formalism, is reached with the minimization
of a free energy model based on the quadratic functional (\ref{i.ug2}) for 
the system potential energy and the use of the CPDF $g_{vv'}(\omega)$.
%

\section{Poisson ensembles} \label{s.pe}

In this section we deduce an exact expression of $g_{vv'}(\omega)$ for an
infinite system composed of independent and randomly distributed points,
the so-called Poisson ensemble. This system is equivalent to the thermodynamic 
limit of a fluid with non-interacting particles (the perfect or ideal gas,
sometimes called Poisson fluid in the mathematics literature).
Poisson ensemble serves as a prototype of more general point systems. 

The nearest neighbor distribution for random configurations of non-interacting 
points in three dimension was derived by Hertz \cite{He09}.
In the present notation, the distribution $n_v$ for Poisson ensembles in 
arbitrary dimension is given by
\beq \label{e.nv}
n_v=n^2e^{-nv}.
\eeq
This distribution can also be derived from the TSP method \cite{Ro05}.
Indeed, (\ref{e.nv}) is the minimum Helmholtz energy solution 
[Eq. (\ref{e.Nv2}) with $\phi_v=0$] for the distribution of available 
volume with the constrains of particle conservation (\ref{c.N}) and space 
normalization (\ref{c.V}) in the thermodynamic limit.

For the evaluation of $g_{vv'}(\omega)$, we use two basic properties 
which characterize to a Poisson point process in arbitrary 
dimension and with average density $n$:
\beqn \label{e.pA}
&&\text{($A$) the probability that a given volume $\Omega$ is found}\cr
&&\text{empty of points, except by a prescribed and}\cr
&&\text{finite number of them, is $e^{-n\Omega}$,}
\eeqn
\beqn \label{e.pB}
&& \text{($B$) the probability of finding a point within an} \cr
&& \text{infinitesimal volume $d\Omega$ is $n d\Omega$.}
\eeqn
%

%
\begin{figure}
\scalebox{0.40}{\includegraphics{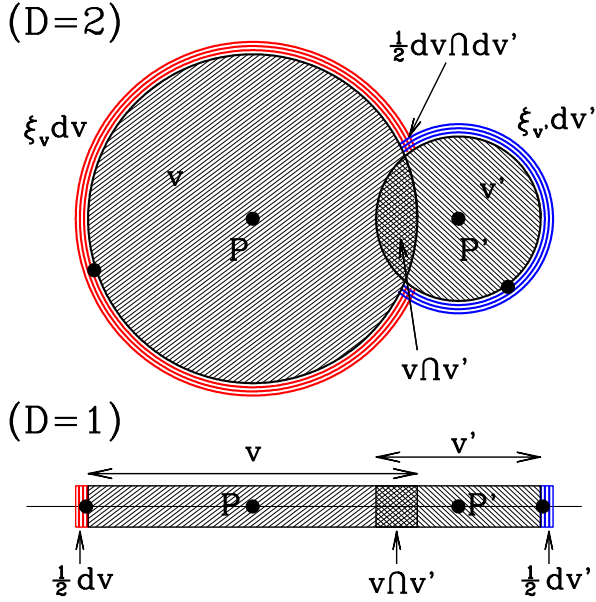}}
\caption{\label{f.circ1} (Color online)
Pairs (v,v') showing various geometrical quantities 
involved in the evaluation of the CPDF $g_{vv'}(\omega)$ for Poisson 
ensembles in one and two dimensions.
}
\end{figure}
\begin{figure}
\scalebox{0.40}{\includegraphics{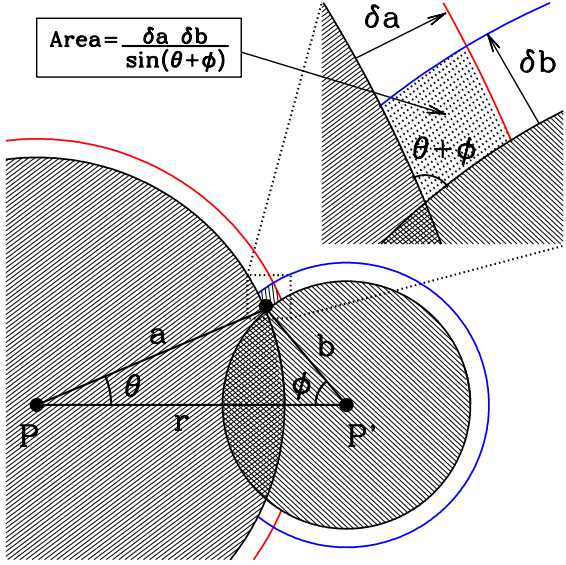}}
\caption{\label{f.circ2} (Color online)
Shared neighbor configuration for $D\ge 2$.}
\end{figure}

We now apply these properties in the evaluation of the occurrence probability
of pairs ($v$,$v'$) in classes (i-iv). We call $\de p$ the probability for the
occurrence of a pair ($v$,$v'$) with the second point (say $P'$), which has
an AV between $v'$ and $v'+\de v'$, lying between the surfaces of volumes 
$\omega$ and $\omega+\de \omega$ centered in the reference point (say $P$),
which has an AV between $v$ and $v+\de v$. By definition of $g_{vv'}(\omega)$,
Eq. (\ref{gvv.def}), it is evident that the probability $\de p$ can be expressed
as follows
\beq \label{e.dpg}
\de p= n^{-1}n_v n_{v'} g^{(\nu)}_{vv'}(\omega)\de v\de v'\de \omega.
\eeq
On the other hand, in the case of configurations class (iv) and according to 
(\ref{e.pA}) and (\ref{e.pB}), the probability $\de p$ equals the product of 
the following four probabilities (see Fig. \ref{f.circ1}):\\
\\
($a$) the probability ($n\de\omega$) that the point $P'$ ($P$) lies between the 
surfaces of $\omega$ and $\omega+\de\omega$ centered in $P$ ($P'$),\\
\\
($b$) the probability ($n\xi_v\de v$) of finding a point between the surfaces 
of $v$ and $v+\delta v$ centered at $P$, where $\xi_v$ is the fraction of 
$\de v$ that is not shared with $v'$,
\beq
\xi_{v}\equiv 1 -\frac{\de v\cap v'}{\de v},
\eeq
\\
($c$) the probability ($n\xi_{v'}\de v'$) of finding a point between the 
surfaces of $v'$ and $v'+\de v'$ centered at $P'$, $\xi_{v'}$ being the 
fraction of $\de v'$ that is not shared with $v$,
\beq
\xi_{v'}\equiv 1 -\frac{\de v'\cap v}{\de v'},
\eeq
\\
($d$) the probability ($e^{-n v\cup v'}$) that no points, apart from $P$ and 
$P'$, lie in the union volume $v\cup v'$.\\
\\
For notational simplicity, explicit $\omega$ and $v'$ ($v$) dependence has 
been omitted for $\xi_v$ ($\xi_{v'}$).
In ($b$) and ($c$) we have used the fact that, due to the meaning of available
volume, there is no point into portions of $\de v$ and $\de v'$ which 
overlap with $v'$ and $v$, respectively.
Thus, for pairs in class (iv) we obtain 
\beq \label{e.dp4}
\de p = e^{-n v\cup v'} (n \xi_{v}\de v) (n\xi_v' \de v') (n\delta \omega)
\Theta(\omega-v_+),
\eeq
where $\Theta(x)$ is the usual Heaviside step function, which here takes into 
account that pair configurations with $\omega<\max \{v,v'\}$ are forbidden
[Eq. (\ref{e.giv0})].
Therefore, using (\ref{e.nv}) and that $v\cup v'=v+v'-v\cap v'$,
it follows from (\ref{e.dpg}) and (\ref{e.dp4}) that 
\beq \label{p.giv}
 g_{vv'}^{(iv)}(\omega) = \xi_{v} \xi_{v'} e^{n v\cap v'} \Theta(\omega-v_+).
\eeq
A similar procedure applies inmediately to pairs ($v$,$v'$) in classes (i-iii).
In the case (iii) we must replace the assumptions ($b$) and ($c$) by the 
probability ($n \de v\cap \de v'$) of finding the shared nearest neighbor 
inside of intersection volume ($\de v \cap \de v'$) of the spherical shells 
$\de v$ and $\de v'$ surrounding $P$ and $P'$, respectively 
(Fig. \ref{f.circ2}). Then
\beq \label{p.giii}
 g_{vv'}^{(iii)}(\omega) = \frac{\eta_{vv'}}n e^{n v\cap v'}
 \Theta(\omega-v_+),
\eeq
where $\eta_{vv'}$ represents the ratio between the volume of 
intersection and the volume product of both shells,
\beq \label{eta_def}
\eta_{vv'}\equiv \frac{\delta v \cap \delta v'}{\delta v \delta v'}
\eeq
(the $\omega$ dependence of $\eta_{vv'}$ is omitted for brevity).
For pairs of class (i), i.e. mutual neighbors, the probabilities ($b$) 
and ($c$) are just considered by ($a$) and must be replaced by Dirac delta 
distributions as expressed by Eq. (\ref{e.gi}). Then
\beq \label{p.gi}
g_{vv'}^{(i)}(\omega)= \frac 1{n^2}\delta(v-v')\delta(\omega-v) e^{n \zeta v},
\eeq
where $\zeta$ is the fraction of the overlap volume ($v\cap v'/v$) of two 
spheres with equivalent size and whose centers are separated by a distance 
equal to their radii (notice that $\zeta$ depends exclusively on the 
dimensionality $D$).  Some $\zeta$ values are given in Table \ref{t.2} below.
Similarly, in the case of pairs of class (ii), the probability of finding a 
nearest neighbor as given by either ($b$) or ($c$) is just included in the 
probability ($a$) and, according to (\ref{e.gii}), we have
\beq \label{p.gii}
g_{vv'}^{(ii)}(\omega)= \frac{\xi_{v_-}}n\delta(\omega-v_+) e^{n v\cap v'}.
\eeq
with 
\beq
v_-\equiv \min(v,v').
\eeq
%

Substitution of expressions (\ref{p.giv}), (\ref{p.giii}), (\ref{p.gii}) and
(\ref{p.gi}) into (\ref{e.gvv}) yields the CPDF for Poisson ensembles 
\beqn \label{gvv_poisson}
g_{vv'}(\omega)&=&e^{n v\cap v'}\left[\left(\frac{\delta(v-v')}{n^2}
    +\frac{\xi_{v_-}}{n}\right)\delta(\omega-v_+) \right.\cr
    &+&\left. \left( \frac{\eta_{vv'}}n +\xi_v \xi_{v'} \right)
    \Theta(\omega-v_+) \right].
\eeqn
The cross-section $v\cap v'$ between $v$ and $v'$ for one-dimensional pairs 
($v,v'$) is easily deduced to be $(v+v'-\omega)/2$ at $\omega\le v+v'$ and 
zero otherwise.
Results for $D=2$ and $D=3$ can be also deduced by simple geometry.
Specifically, one finds that \cite{RS11}
\beq \label{e.vcapv}
v\cap v' = \left\{
\begin{array}{l l}
   \frac 12(v+v'-\omega),  & (D=1) \\
   \frac v\pi \left[\theta -\frac 12\sin(2\theta)\right] &  \\
 + \frac {v'}\pi \left[\phi-\frac 12\sin(2\phi)\right],  & (D=2) \\
   \frac v4  (2-3\cos\theta+\cos^3\theta) & \\
 + \frac {v'}4 (2-3\cos\phi+\cos^3\phi), & (D=3) \\
\end{array}
\right.
\eeq
for $v_+\le \omega\le \omega_{vv'}$ and $v\cap v'=0$ for $\omega >\omega_{vv'}$,
where the angles $\theta$ and $\phi$ are given by ($D\ge 2$)
\beq \label{e.theta}
\cos \theta=\frac{\omega^{2/D}+v^{2/D}-v'^{2/D}}{2(\omega v)^{1/D}},\;
\eeq
\beq \label{e.phi}
\cos \phi=\frac{\omega^{2/D}+v'^{2/D}-v^{2/D}}{2(\omega v')^{1/D}}.
\eeq
The quantities $\xi_{v}$, $\xi_{v'}$ $\eta_{vv'}$ in (\ref{gvv_poisson}) 
can be evaluated in terms of elementary functions.
For $\omega > \omega_{vv'}$, with $\omega_{vv'}$ given by Eq. (\ref{wvv}), 
we have $\xi_v =\xi_{v'}=1$, $\eta_{vv'}=0$, $v\cap v'=0$, and therefore 
\beq \label{gp1} 
g_{vv'}(\omega)=g^{(iv)}_{vv'}(\omega)=1,\quad  \omega > \omega_{vv'}.
\eeq
%
For $v_+\le \omega \le \omega_{vv'}$, it can be shown easily (with help of 
Figs. \ref{f.circ1} and \ref{f.circ2}) that
\beq \label{e.xiv}
\xi_{v} = \left\{
\begin{array}{l }
   \frac 12,               \\ 
   1-\theta/\pi,      \\
   \frac 12(1+\cos \theta), \\
\end{array}
\right.
\xi_{v'} = \left\{
\begin{array}{l l}
   \frac 12,              & (D=1) \\
   1-\phi/\pi,       & (D=2) \\
   \frac 12(1+\cos \phi),  & (D=3) \\
\end{array}
\right.
\eeq
and
\beq \label{e.eta}
\eta_{vv'} = \left\{
\begin{array}{l l}
 \frac 12 \delta(\omega-v-v'),  & (D=1)  \\
 (2\pi \sqrt{vv'} \sin[\theta+\phi])^{-1},  & (D=2)  \\
 \sin\theta[6 v^{1/3}v'^{2/3}\sin(\theta+\phi)]^{-1}.  & (D=3)  \\
\end{array}
\right.
\eeq
Some guidelines for the derivation of $\eta_{vv'}$ are given in Appendix
\ref{a.eta}. Although $\eta_{vv'}$ was written in a particular way for $D=3$, 
it is actually symmetric with respect to $v$ and $v'$.
Notice also that the prefactor $1/2$ in (\ref{e.xiv}) and (\ref{e.eta}) for 
$D=1$ is related to the ``edge surface'' of the one-dimensional ``volumes'' 
$v$ and $v'$ (see Fig. \ref{f.circ1}).

\subsection{Explicit expressions for $D=1$ and $D\rightarrow \infty$}
\label{s.D1i}

Substantial simplifications in Eq. (\ref{gvv_poisson}) occur in the limits 
of one and infinitely many space dimensions. In particular, the combination 
of (\ref{gvv_poisson}) with results (\ref{e.vcapv}), (\ref{e.xiv}) and 
(\ref{e.eta}) at $D=1$, gives the CPDF for one dimensional Poisson systems
\beqn \label{gvv_1D}
g_{vv'}(\omega)&=&\left[\frac{e^{(nv)/2}}{n^2} \delta(v-v')
    +\frac{e^{(nv_-)/2}}{2n}\right] \delta(\omega-v_+) \cr
    &+&\frac{e^{n(v+v'-\omega)/2}}{4} \Theta(v+v'-\omega)\Theta(\omega-v_+) \cr
    &+&\frac 1{2n} \delta(\omega-v-v') + \Theta(\omega-v-v').
\eeqn
On the other hand, as the spatial dimension tends to infinite the intersection
volume of two off-center hyperspheres ($v\cap v'$), the overlap space between 
two off-center hyperspherical shells ($\de v \cap \de v'$), and the cross 
section between a hyperspherical shell and a hypersphere ($\de v \cap v'$) no 
centered, all go to zero. 
Consequently, $\xi_v=\xi_{v'}\rightarrow 1$ and $\eta_{vv'}\rightarrow 0$ 
everywhere as $D\rightarrow \infty$, and therefore (\ref{gvv_poisson}) 
reduces to
\beq
g_{vv'}(\omega)=\left[\frac{1}{n^2} \delta(v-v')
    +\frac{1}{n}\right] \delta(\omega-v_+) + \Theta(\omega-v_+).
\eeq
One of the main differences between the CPDF of Poisson ensembles at one and 
infinite space dimensions concerns the shared neighbor contribution.
For $D=1$, $g^{iii}_{vv'}(\omega)$ is a singular function that corresponds 
to a single point of contact between $v$ and $v'$, while that at $D=\infty$
the probability density of finding a pair sharing the first neighbor tends 
to zero everywhere because $\eta_{vv'}\rightarrow 0$.

%
\begin{figure}
\scalebox{0.40}{\includegraphics{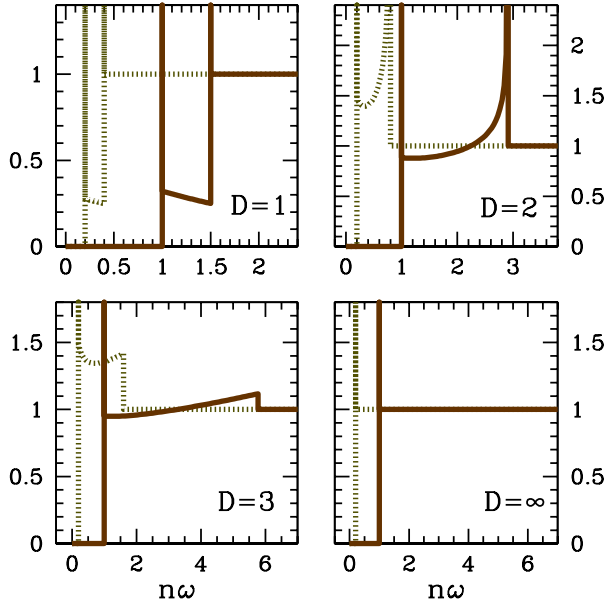}}
\caption{\label{f.gvv} (Color online)
Evaluations of $g_{vv'}(\omega)$ at dimensions $D=1,2,3,\infty$
for $(nv,nv')=(0.2,0.2)$ (dotted lines) and $(nv,nv')=(1,0.5)$ (solid lines).
}
\end{figure}
%
%
\begin{figure}
\scalebox{0.40}{\includegraphics{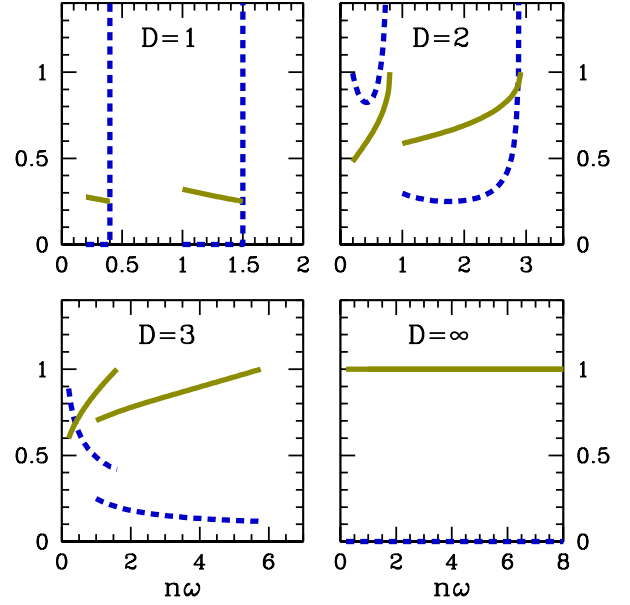}}
\caption{\label{f.gvv34} (Color online)
Contributions $g^{(iii)}_{vv'}(\omega)$ and $g^{(iv)}(\omega)$ (dashed and 
solid lines, respectively) to $g_{vv'}(\omega)$ at $v_+<\omega<\omega_{vv'}$,
for $D=1,2,3,\infty$, and for the two ($v,v'$) conditions shown in 
Fig. \ref{f.gvv}.
}
\end{figure}
%

\subsection{Discussion}

In Fig. \ref{f.gvv} we depict evaluations of $g_{vv'}(\omega)$  based on
Eq. (\ref{gvv_poisson}) for $D=1,2,3$, and $\infty$, at $(nv,nv')=(0.2,0.2)$ 
and $(nv,nv')=(1,0.5)$ (dotted and solid lines, respectively). For the same 
conditions, Fig. \ref{f.gvv34} shows the contributions to $g_{vv'}(\omega)$ 
origined by pairs ($v,v'$) in classes (iii) and (iv), as given by Eqs. 
(\ref{p.giii}) and (\ref{p.giv}). The contribution which shows the largest 
variations is $g^{(iii)}_{vv'}(\omega)$, a direct consequence of the complex 
dependence of the overlap shell volume $\eta_{vv'}$ with the parameters $v$, 
$v'$ and $\omega$.

As observed in Fig. \ref{f.gvv}, $g_{vv'}(\omega)$ vanishes for all dimension 
at $\omega<v_+$, which means that there are not very close pairs $(v,v')$ as
a consequence of the exclusion zones imposed by the AV $v$ and $v'$.
The main striking features on the plots are the strong correlations between 
pairs ($v,v'$) at $\omega=v_+$ for all dimensions and at $\omega=\omega_{vv'}$ 
for $D=1$ and $2$. For, respectively, $D=1,2$ and $3$, 
$\omega_{vv'}=0.4$, $0.8$ and $1.6$ in the case of pairs $(nv,nv')=(0.2,0.2)$,
and $\omega_{vv'}=1.5$, $2.914$ and $5.771$ in the case $(nv,nv')=(1,0.5)$.

The singularity at $\omega=v_+$, denoted by vertical lines in Fig. \ref{f.gvv}, 
is produced by pairs where at least one of the two points is the nearest 
neighbor of the other one, i.e., it has origin in pairs of type (ii), together 
with pairs of type (i) for the example with $v=v'$ (mutual nearest neighbors). 
The second strong correlation originated at one and two dimensions is due to 
the shared nearest neighbor configuration, i.e., pairs in class (iii).
For $D=1$, the singularity introduced by the $\delta(\omega-v-v')$ function 
in Eq. (\ref{gvv_1D}), seen as vertical lines in Fig. \ref{f.gvv34}, 
is a consequence of the fact that the shared neighbor must be located in the 
intersection of the outer edges of $v$ and $v'$, 
which for $D=1$ corresponds to a single point in space.
In the case of two-dimensional Poisson ensembles, the scaled overlap volume
$\eta_{vv'}$ gives nonzero values in the range $v_+ <\omega<\omega_{vv'}$ and 
has a singularity at $\omega=\omega_{vv'}$, which is reflected in a dramatic 
rise of $g_{vv'}(\omega)$ (Fig. \ref{f.gvv}, $D=2$).

At $D=3$, the overlaping $\eta_{vv'}$ of the outer edges of $v$ and $v'$ 
is less sensitive to variation of $\omega$ and therefore the contribution 
$g^{(iii)}_{vv'}(\omega)$ remains bounded for all $\omega$ 
(Fig. \ref{f.gvv34}). In this dimension, the discontinuity in 
$g_{vv'}(\omega)$ at $\omega=\omega_{vv'}$ (Fig. \ref{f.gvv}) is yielded 
by the sudden fall of the occurrence of shared neighbor pairs.
For $D=\infty$, the contribution of shared nearest neighbor configurations
is everywhere negligible (dashed line in Fig. \ref{f.gvv34}) and that from
pairs of class (iv) is uniform and equals unity (solid line), the total 
CPDF resulting flat at $\omega>v_+$ (Fig. \ref{f.gvv}).

\section{The CPDF $g_v(\omega)$ at Poisson ensembles}

An expression for $g_v(\omega)$ corresponding to Poisson ensembles can be 
obtained in a similar, but simpler and more straightforward way than that 
previously used for $g_{vv'}(\omega)$ (its derivation is left to the reader).
This CPDF was previously introduced and analyzed in \cite{Ro05} and reads
\beqn \label{gv_poisson}
g_{v}(\omega)=\frac{1}{n}\delta(\omega-v) +\Theta(\omega-v).
\eeqn
Here, the Dirac delta function accounts for the nearest neighbor contribution 
of a reference point with AV $v$, and the Heaviside function describes 
the probability density of finding points (other than the nearest one) lying
in the surface of $\omega>v$ centered in the reference one.
It should be noted that (\ref{gv_poisson}) could be derived from (\ref{gvgvv})
and (\ref{gvv_poisson}), but this procedure involves some analytically 
intractable integrals for $D\ge2$.

On the other hand, substitution of (\ref{gv_poisson}) into (\ref{ggv}) 
provides the expected PDF for Poisson ensembles, i.e.,
\beqn \label{g_poisson}
g(\omega)=1.
\eeqn
This result reflects the well-known fact that no correlation exists between 
pairs of arbitrary points at Poisson ensembles. Clearly the departures
from unity in $g_v(\omega)$ and $g_{vv'}(\omega)$ are due to that the available 
volumes $v$ and $v'$ introduce correlations between pairs at short distances, 
which correspond to $\omega\le v$ and $\omega\le \omega_{vv'}$ for 
$g_v(\omega)$ and $g_{vv'}(\omega)$, respectively.

\subsection{Relative importance of the different contributions
to the CPDF $g_v(\omega)$}

The results obtained in the derivation of $g_{vv'}(\omega)$ can be used to 
evaluate various statistical neighboring functions. 
In particular, expressions (\ref{ggv}) and (\ref{gvgvv}) can be applied to
each class ($\nu=$ i, ii, iii, iv) of pairs ($v,v'$),
\beq \label{ggv_nu1}
g^{(\nu)}_v(\omega)=\frac 1n\int_0^\infty n_{v'}g^{(\nu)}_{vv'}(\omega)dv',
\eeq
\beq \label{ggv_nu2}
g^{(\nu)}(\omega) = \frac 1n\int_0^\infty n_{v}g^{(\nu)}_{v}(\omega)dv.
\eeq
The quantities in Eqs. (\ref{ggv_nu1}) and (\ref{ggv_nu2}) express partial 
probability densities that describe the spatial distributions of specific 
pair configurations.
For mutual neighbor pairs, from (\ref{p.gi}), (\ref{ggv_nu1}), and 
(\ref{ggv_nu2}) we obtain
\beq
g^{(i)}_v(\omega)=\frac 1n e^{(\zeta-1)nv} \delta(\omega-v),
\eeq
\beq  \label{gi}
g^{(i)}(\omega) = e^{(\zeta-2)n\omega}. 
\eeq
Here, it is useful to separate pairs in class (ii) into subclasses 
(ii$a$) and (ii$b$). Pairs ($v,v'$) where the point with AV $v'$ is the 
(non mutual) nearest neighbor of the point with AV $v$ belong to the subclass 
(ii$a$), otherwise they belong to the subclass (ii$b$). 
Then, Eq. (\ref{ggv_nu1}) yields
\beq \label{gii_v}
g^{(ii)}_v(\omega)= g^{(iia)}_v(\omega)+g^{(iib)}_v(\omega)
\eeq
with
\beq \label{giia_v}
g^{(iia)}_v(\omega)=
\delta(\omega-v) \int_0^v \xi_{v'} e^{-nC_{v'}}dv',
\eeq
%
\begin{figure}
\scalebox{0.40}{\includegraphics{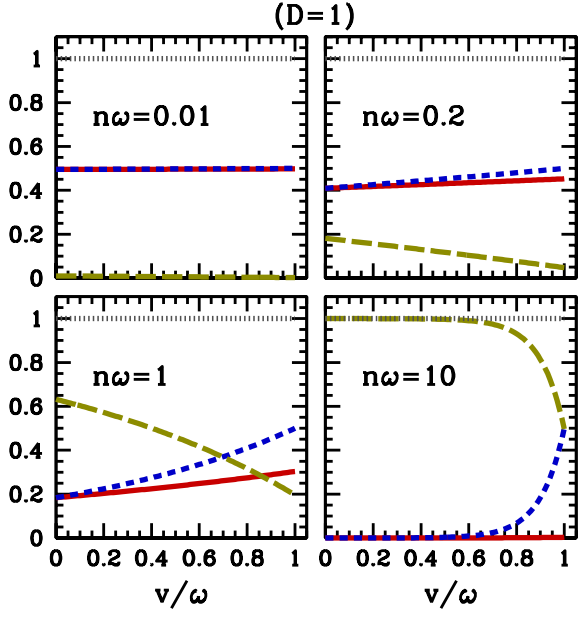}}
\caption{\label{f.gvD1} (Color online)
Contributions due to (ii$b$), (iii) and (iv) pair configurations (solid, short 
dashed and long dashed lines, respectively) to the CPDF $g_v(\omega)$ as a
function of $v$, for one dimensional Poisson ensembles and different distances 
as measured by $\omega$ (values $n\omega$ indicated on the plots).
The total CPDF is shown by the dotted line.
}
\end{figure}
%
%
\begin{figure}
\scalebox{0.40}{\includegraphics{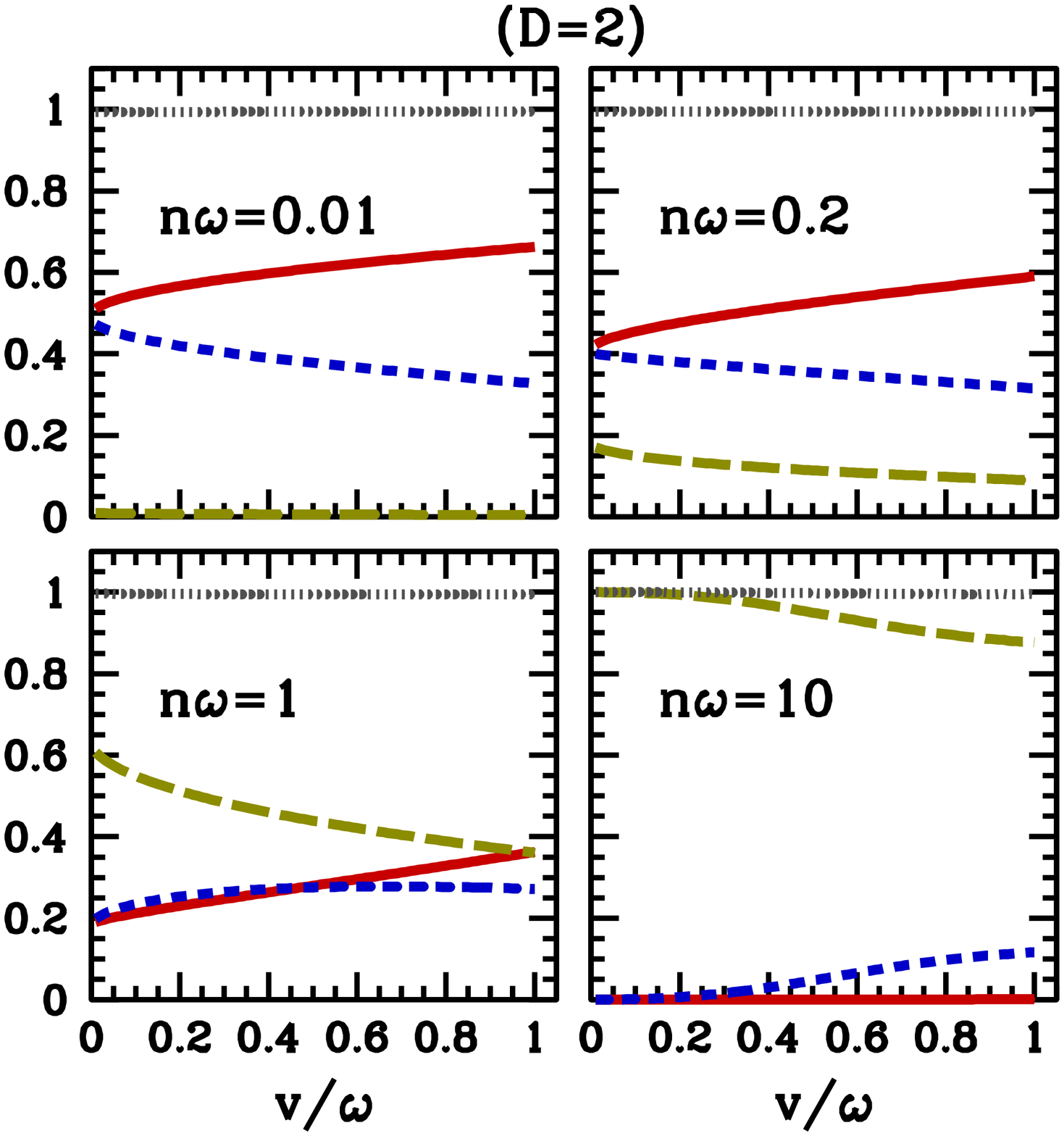}}
\caption{\label{f.gvD2} (Color online) Same as Fig. \ref{f.gvD1}, for $D=2$.
}
\end{figure}
%
%
\begin{figure}
\scalebox{0.40}{\includegraphics{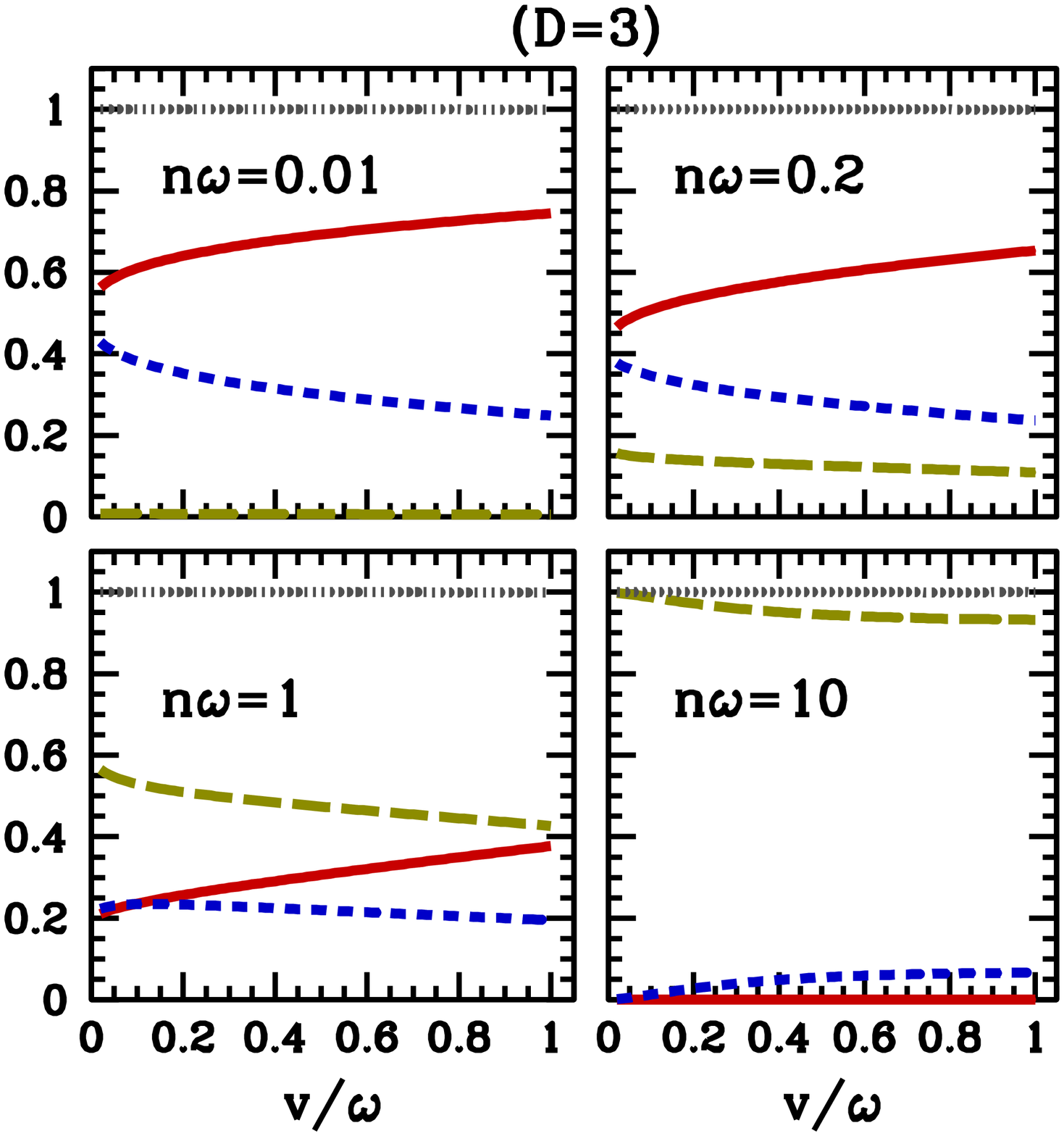}}
\caption{\label{f.gvD3} (Color online) Same as Fig. \ref{f.gvD1}, for $D=3$.
}
\end{figure}
%
%
\begin{figure}
\scalebox{0.40}{\includegraphics{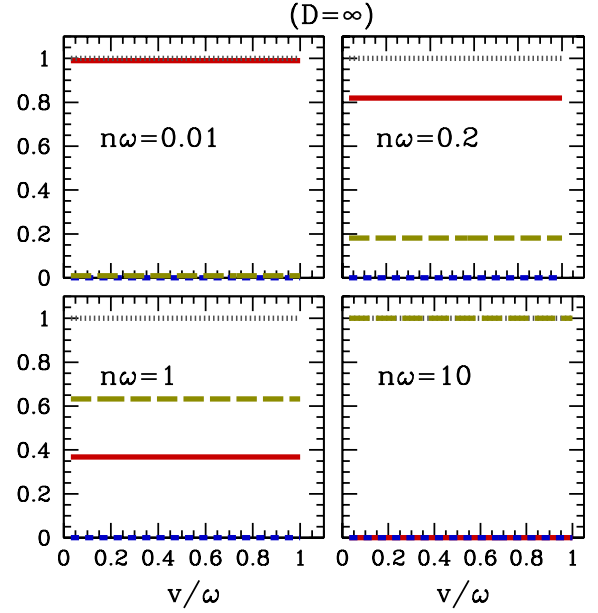}}
\caption{\label{f.gvDi} (Color online) 
Same as Fig. \ref{f.gvD1}, for $D=\infty$.
}
\end{figure}
%
and
\beq \label{giib_v}
g^{(iib)}_v(\omega)=
     \left[\xi_v e^{-nC_{v'}} \right]_{v'=\omega}\Theta(\omega-v) ,
\eeq
where $C_{v'}=v'-v\cap v'$ is the volume of a crescent in the intersection of 
$v$ and $v'$. 
The term $g^{(iia)}_v(\omega)$ is the probability density of
finding a non-mutual, first neighbor in the surface of $\omega$ centered in 
a reference point with AV $v$. 
The term $g^{(iib)}_v(\omega)$ is the contribution to $g_v(\omega)$ due to 
pairs where the reference point with AV $v$ is the non-mutual nearest 
neighbor of the other one. 
Note that the Dirac delta term in (\ref{gv_poisson}) arises from contributions 
due to $g^{(i)}_v(\omega)$ and $g^{(iia)}_v(\omega)$ through (\ref{gvgvv}), 
which then implies that the integral in (\ref{giia_v}) is given by
\beq \label{integral1}
\int_0^v \xi_{v'} e^{-nC_{v'}}dv'= \frac 1n \left[1-e^{(\zeta-1)nv}\right].
\eeq
The contribution to the PDF $g(\omega)$ due to pairs in subclasses (iia) and 
(iib) may be found from (\ref{ggv_nu2}), (\ref{giia_v}) and (\ref{giib_v}).  
Thus, using (\ref{integral1}), one obtains
\beq \label{gii_a}
g^{(iia)}(\omega)= e^{-n\omega}-e^{(\zeta-2)n\omega}.
\eeq
By symmetry, pairs in subclasses (ii$a$) and (ii$b$) have equivalent PDFs, 
i.e., $g^{(iib)}(\omega)= g^{(iia)}(\omega)$, which implies
\beq \label{gii_b}
 g^{(iib)}(\omega)=
    \int_0^\omega n\left[\xi_v e^{-nv\cup v'} \right]_{v'=\omega}dv
    = e^{-n\omega}-e^{(\zeta-2)n\omega}.
\eeq

A similar analysis applies to pairs in classes (iii) and (iv) but their 
contributions to $g_v(\omega)$ and $g(\omega)$ cannot be determined 
analytically except for $D=1$ and $D=\infty$. 
In general, these functions are expressed as integral forms,
\beq \label{giii_v}
g_v^{(iii)}(\omega)= \int_0^\omega \eta_{vv'} e^{-nC_{v'}}dv'\Theta(\omega-v),
\eeq
\beq \label{giv_v}
g_v^{(iv)}(\omega)=n \int_0^\omega \xi_v\xi_{v'} e^{-nC_{v'}}dv'
        \Theta(\omega-v),
\eeq
\beq \label{giii}
g^{(iii)}(\omega)=n\int_0^\omega \int_0^\omega \eta_{vv'} e^{-n v\cup v'}dv'dv,
\eeq
\beq \label{giv}
g^{(iv)}(\omega)=n^2 \int_0^\omega \int_0^\omega \xi_v\xi_{v'} 
                 e^{-nv\cup v'}dv'dv.
\eeq
Again, by virtue of (\ref{gvgvv}), the Heaviside term in (\ref{gv_poisson})
comes from (\ref{giib_v}), (\ref{giii_v}) and (\ref{giv_v}), therefore
\beq \label{gviii_iv}
g_v^{(iii)}(\omega) + g_v^{(iv)}(\omega) = 
 \left[1-\left(\xi_v e^{-nC_{v'}} \right)_{v'=\omega}\right] \Theta(\omega-v).
\eeq
which implies
\beq \label{integral2}
\int_0^\omega\left( \eta_{vv'} +n\xi_v\xi_{v'}\right) e^{-nC_{v'}}dv' 
 = 1-\left[\xi_v e^{-nC_{v'}} \right]_{v'=\omega}.
\eeq
Similarly, with (\ref{gii_b}) and (\ref{gviii_iv}) we obtain
\beq \label{giii_iv}
g^{(iii)}(\omega) + g^{(iv)}(\omega) = 1-2e^{-n\omega} +e^{(\zeta-2)n\omega}.
\eeq
Explicit mathematical formulas for $g^{(\nu)}_v(\omega)$ and 
$g^{(\nu)}(\omega)$ functions at $D=1$ and $D=\infty$ have been collected 
in Appendixes \ref{a.D1} and \ref{a.Dinf}.

The explicit expression for $g^{(i)}(\omega)$ at $D=3$ [as given by Eq. 
(\ref{gi})] was previously obtained by Larsen \& Stratt \cite{LS98}. 
To the best of our knowledge, all other results derived here are new.

\subsection{Discussion}

Figures \ref{f.gvD1}, \ref{f.gvD2}, \ref{f.gvD3} and \ref{f.gvDi} depict the 
partial probability densities $g^{(iib)}_v(\omega)$ (solid lines), 
$g^{(iii)}_v(\omega)$ (dashed lines) and $g^{(iv)}_v(\omega)$ (long dashed 
lines) as functions of the AV $v$ ($0\le v< \omega$), 
for different volume $\omega$ and dimensions $D$. These evaluations come from 
Eqs. (\ref{giib_v}), (\ref{giii_v}), and (\ref{giv_v}).
It is worth to notice that the sum of these three contributions gives the 
unity (dotted lines in the figures), which is actually the value of the 
total CPDF $g_v(\omega)$ for $v<\omega$ [Eq. (\ref{gv_poisson})].

At finite dimension and short pair separations ($n\omega\ll 1$), the CPDF 
$g_v(\omega)$ is almost entirely composed of contributions due to pairs in 
classes (ii$b$) and (iii), i.e., the reference point with AV $v$ is the 
nearest neighbor of the other point at $\omega$ or both have a shared nearest 
neighbor. In the one dimensional system (Fig. \ref{f.gvD1}), 
shared neighbor configurations (short dashed lines) are somewhat more likely 
than those in class (ii$b$), while for $D>1$ the major contribution 
arises from pair of class (ii$b$) (solid lines, Figs. \ref{f.gvD2} and 
\ref{f.gvD3}).
Pairs sharing the nearest neighbor becomes highly unlikely at high space
dimensionalities for any pair separation (Fig. \ref{f.gvDi}).
As expected, class (iv) (long dashed lines) is the dominant pair 
configuration at any $D$ for great distances ($n\omega>1$) from the reference 
point.

At finite dimension, as the available volume $v$ of the reference point 
increases with fixed $\omega$, the surface fraction $\xi_v$ in Eq. 
(\ref{giib_v}) increases and the exclusion volume $C_{v'=\omega}$ decreases. 
Consequently, the probability of finding pairs at class (ii$b$) increases 
monotonically with $v$.
On the contrary, the chances of finding four-point configurations (pairs at 
class iv) reduce with increasing $v$ (or decreasing $\omega$) because
the surface fraction $\xi_{v'}$ at the integrand of Eq. (\ref{giv_v}) takes 
on average lower values (which are not fully compensated by the increasing 
of the other terms in the integrand).
The higher the Euclidean space dimensionality, the weaker the $v$
dependence of the geometrical quantities ($\xi_v$, $\eta_{vv'}$, $C_{v'}$),
and the partial probability densities $g^{(\nu)}_v(\omega)$ analyzed 
here become independent of $v$ at the limit $D\rightarrow\infty$ (Fig. 
\ref{f.gvDi}).
%
\begin{figure}
\scalebox{0.40}{\includegraphics{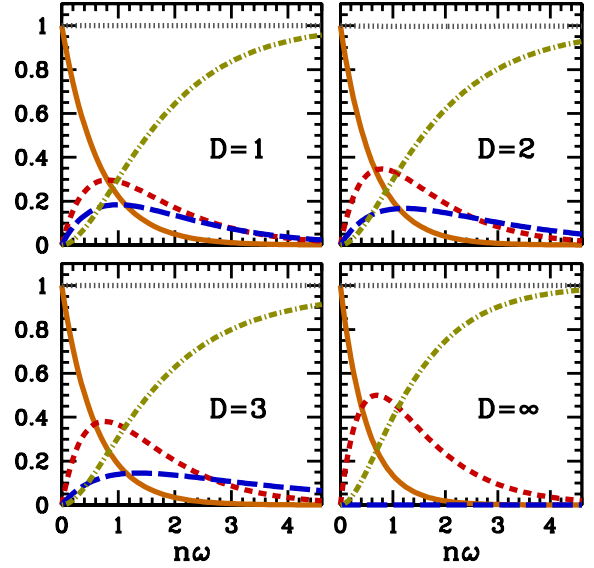}}
\caption{\label{f.g} (Color online)
Contributions $g^{(\nu)}(\omega)$ due to $\nu=$i, ii, iii and iv pair 
configurations (solid, short dashed, long dashed and point-dashed lines, 
respectively) to the PDF $g(\omega)$ vs the volume $\omega$, 
for Poisson ensembles at dimensions $D=1,2,3$ and $\infty$.
The total PDF is shown by a dotted line.
}
\end{figure}
%
\subsection{Relative importance of the different contributions
to the PDF $g(\omega)=1$}

Figure \ref{f.g} illustrates the behavior of the partial probability 
densities $g^{(\nu)}(\omega)$ given by Eqs. (\ref{gi}), (\ref{gii_a}),
(\ref{gii_b}), (\ref{giii}) and (\ref{giv}). Note that $g^{(ii)}(\omega)$ is 
given by the sum of (\ref{gii_a}) and (\ref{gii_b}).
The term $g^{(i)}(\omega)$ dominates the regime of low volume ($n\omega\ll 1$) 
at any $D$, i.e., most points with a first neighbor at very short distance 
(lower than the mean separation between points in the ensemble) form
pairs of mutual nearest neighbors.
At intermediate distances from a reference point (which correspond to 
$n\omega\approx 1$), there is a relatively narrow $\omega$ range where pairs 
in class (ii) are the main contribution to $g(w)$ for any $D$. 
As the distance from the reference point increases ($n\omega >1$), 
pairs in class (iv) become the dominant contribution to $g(\omega)$.
Indeed, $g^{(iv)}(\omega)$ is a monotonically increasing function of
$\omega$ and tends asymptotically to unity, while the remaining contributions 
go to zero for $\omega$ large enough.
The term $g^{(iii)}(\omega)$ corresponding to pairs with the shared nearest 
neighbor is never dominant and has a maximum contribution of $18.39\%$, 
$16.72\%$ and $14.54\%$ at $n\omega=1$, $1.226$ and $1.370$ for $D=1,2$ and 3, 
respectively. This maximum value 
gradually decreases and shifts to larger $\omega$ with increasing dimension 
$D$. The converse behavior is followed by $g^{(ii)}(\omega)$. Indeed, the
maximum probability for the occurrence of pairs where one and only one of 
the two points is the nearest neighbor of the other occurs at 
$n\omega=0.811$ ($29.63\%$), $0.781$ ($34.67\%$),  $0.761$ ($38.06\%$), 
and $0.694$ ($50\%$) for $D=1,2,3$ and $\infty$.
\begin{table}
\vspace{-0.5cm}
\caption{Average fractions of pairs in various classes} \label{t.2}
\vspace{-0.cm}
\begin{ruledtabular}
\begin{tabular}{llllccc}
$D$ & $\alpha$ & $\zeta$ &  $f^{(i)}$ & $f^{(iia)}=f^{(iib)}$ & $f^{(iii)}$ 
& $f^{(iii)[*]}$ \\
\hline
1 & 2        & $1/2$                       & 0.667 & 0.333 & 0.500 & 0.50 \\
2 & $\pi$    & $\frac{4\pi-3^{3/2}}{6\pi}$ & 0.622 & 0.378 & 0.618 & 0.63\\
3 & $4\pi/3$ & $5/16$                      & 0.593 & 0.407 & 0.709 & 0.71 \\
$\infty$ &  0  & 0                         & 0.500 & 0.500 &   -   & 1.00 \\
\end{tabular}
$^{*}$ Values listed in Ref. \cite{Sc86}
\end{ruledtabular}
\end{table}
\subsection{Pair fractions and mean separations}

Finally we mention that one can use $g_{vv'}(\omega)$ to obtain other useful
measures of the structure of Poisson ensembles or many-body systems in general.
For examples, the fraction $f^{(\nu)}$ of points in classes $\nu=$i, ii and 
iii, averaged over the whole ensemble, and the mean pair separation 
$\la r\ra^{(\nu)}$ in each class. They are defined as
\beq \label{e.fm}
f^{(\nu)} = \int_0^\infty n g^{(\nu)}(\omega)d\omega,
\eeq
and
\beq \label{e.rm}
\la r\ra ^{(\nu)} = \frac 1{f^{(\nu)}}
\int_0^\infty n r(\omega) g^{(\nu)}(\omega)d\omega,
\eeq
with $r(\omega)=(\omega/\alpha)^{1/D}$. 
Results for pairs in classes (i) and (ii) at Poisson ensembles are 
found analytically using Eqs. (\ref{gi}), (\ref{gii_a}) and (\ref{gii_b}). 
Thus
\beq \label{e.fi}
f^{(i)}= \frac 1{2-\zeta}, 
\eeq
\beq \label{e.fii}
f^{(iia)}= f^{(iib)}= \frac {1-\zeta}{2-\zeta},
\eeq
\beq
\la r\ra^{(i)}=\frac {\Gamma\left(1+\frac 1D\right)}
              {\left[n\alpha(2-\zeta)\right]^{1/D}},
\eeq
\beq
\la r\ra^{(ii)}=\frac {\Gamma\left(1+\frac 1D\right)
              \left[(2-\zeta)^{1+1/D}-1\right]}
              {\left[n\alpha(2-\zeta)\right]^{1/D}(1-\zeta)}.
\eeq
Notice that the nearest neighbor of any point is either mutual or non mutual 
one and therefore $f^{(i)}+f^{(iia)}=1$.
Mean values $\la \omega\ra^{(\nu)}$ can be also computed in similar form to 
$\la r\ra^{(\nu)}$,
\beq
\la \omega\ra^{(i)}=\frac {1}{n(2-\zeta)},\quad
\la \omega\ra^{(ii)}= \frac 1n+\la \omega\ra^{(i)}.
\eeq
Clearly, results based on the volume $\omega$ are more concise than those 
upon the distance $r$.
\begin{table}
\vspace{-0.5cm}
\caption{Mean $\omega$-volumes and average distances between points of pairs in 
classes (i), (ii) and (iii). $\la r \ra$ and $\la\omega\ra$ are expressed in 
units of mean distance between points $n^{-1/D}$ and mean volume per point 
$n^{-1}$, respectively.} \label{t.3}
\vspace{-0.cm}
\begin{ruledtabular}
\begin{tabular}{lllllll}
$D$ & $\la r\ra^{(i)}$ & $\la r\ra^{(ii)}$ & $\la r\ra^{(iii)}$ 
 & $\la\omega\ra^{(i)}$ & $\la\omega\ra^{(ii)}$ & $\la\omega\ra^{(iii)}$ \\
\hline
1   &  0.333 &  0.833  &  1.000 & 0.667 & 1.667 & 2.000 \\
2   &  0.394 &  0.674  &  0.880 & 0.622 & 1.622 & 2.797 \\
3   &  0.466 &  0.683  &  0.909 & 0.593 & 1.593 & 3.961 \\
\end{tabular}
\end{ruledtabular}
\end{table}
%

The fraction $f^{(i)}$ of mutual nearest neighbors at Poisson ensembles was
initially studied by Clark \cite{Cl56}, although the exact evaluation 
[equivalent to that given by Eq. (\ref{e.fi})] was obtained by Dacey 
\cite{Da69} and Cox \cite{Co81}. The fraction $f^{(iii)}$ of pairs sharing 
the nearest neighbor has also been computed by Schilling \cite{Sc86}.
All these authors have used techniques different from that adopted in the 
present work, which are based on conditional pair distribution functions. 
Our evaluations of $f^{(\nu)}$, $\la r \ra^{(\nu)}$ and 
$\la \omega\ra^{(\nu)}$, together with values of the parameter $\alpha$ and 
the overlap volume $\zeta$ are listed in Tables \ref{t.2} and \ref{t.3}.
Results corresponding to pairs in class (iii) were evaluated by numerical 
integration of (\ref{e.fm}) and (\ref{e.rm}).
The fractions $f^{(iii)}$ computed here are in agreement with those obtained 
by Schilling \cite{Sc86} (see Table \ref{t.2}). 
Interesting enough, Schilling showed that $f^{(iii)}$ tends to unity as 
$D\rightarrow \infty$. This implies that whereas $g^{(iii)}_{vv'}(\omega)$ 
tends to zero everywhere (see Section \ref{s.D1i}) with $D\rightarrow\infty$, 
its integral over $\omega$ [Eq. (\ref{e.fm})] converges to unity.

\section{ Conclusions} \label{s.co}

A new mathematical descriptor of spatial structures in many-body systems have 
been proposed. Specifically, we have introduced a two-point distribution 
function, denoted by $g_{vv'}(\omega)$ and defined by Eq. (\ref{gvv.def}), 
which describes the probability density of finding pairs of objects with each
object having a nearest neighbor at a certain distance. 
For simplicity in the mathematical framework, distances between objects have 
been substituted by spherical volumes whose radii denote the separation between 
two objets of a pair (volume $\omega$), and between each object and its nearest 
neighbor (volumes $v$ and $v'$). 

The conditional pair distribution function (CPDF) $g_{vv'}(\omega)$ reveals a 
richer structural information than that provided by the conventional PDF 
$g(\omega)$. Indeed, the volumes $v$ and $v'$ introduce correlations 
in the CPDF which are not implicit in $g(\omega)$. Moreover, nearest 
neighbor information from $v$ and $v'$ let us to identify four types of pairs 
(mutual nearest neighbors and neighbor sharing pairs, among others), which 
provide a considerable amount of information concerning the local microscopic 
density and spatial relationships in many-body systems.

General relations were established among $g_{vv'}(\omega)$, the 
standard PDF $g(\omega)$, and $g_v(\omega)$. The function $g_v(\omega)$, 
previously introduced in \cite{Ro05}, is a CPDF in which the nearest 
neighbor information is only considered for one object of each pair.
We have shown that, using appropriate many-body interaction potentials, 
$g_v(\omega)$ and $g_{vv'}(\omega)$ can be used to evaluate the total 
potential energy of a fluid, which reduces to that typically adopted in liquid 
theory with an averaged pairwise potential. It was shown also that the 
potential energy expression based on $g_{vv'}(\omega)$ is the correct one 
to be used in the so-called thermostatistical space partition (TSP) formalism 
\cite{Ro05,RZ06}. 

We have studied the function $g_{vv'}(\omega)$ for Poisson ensembles and 
derived its exact expression in any space dimension. We have obtained 
explicit results for $g_{vv'}(\omega)$ at $D=1,2,3,\infty$, and found simple 
close forms for both $D=1$ and $D=\infty$. 
To our knowledge, all these results are new.
Furthermore, functions $g_{vv'}(\omega)$, $g_{v}(\omega)$ and $g(\omega)$ 
corresponding to Poisson systems have been deconvoluted into contributions 
from the four pair classes. With the exception of the function 
$g^{(i)}(\omega)$ given in \cite{LS98} for mutual nearest neighbors at $D=3$, 
all the partial probability densities obtained in the present study are new.
Our theoretical results should have interesting and 
practical applications in the study of neigborhood structures in a variety 
of areas, such as physical and biological sciences, and sociology.
The extension of present evaluations to non-Poisson processes is open to 
future efforts.
\acknowledgments
The authors are indebted to Professor Andr\'es Santos for helpful suggestions
and comments.
This work has been supported by CONICET (Argentina) through Grant
PIP 112-200801-01474.

\appendix
\section{Covolume of two shells} \label{a.eta}

In this Appendix we give a brief derivation of the scaled overlap volume 
$\eta_{vv'}$ of shells $\de v$ and $\de v'$ surrounding the available volumes 
$v$ and $v'$ [Eq. (\ref{e.eta})]. This quantity is directly related to the 
shared neighbor configuration in pairs ($v,v'$).
The shared nearest neighbor of a pair ($v,v'$) must be located in the 
intersection of the outer edges of $v$ and $v'$. For $D=1$, this intersection
corresponds to a single point in space, which yields the Dirac delta function
at (\ref{e.eta}) with a prefactor $1/2$ because the shared neighbor is located 
in one of the two sides of $v$.
At $D=2$, $\de v\cap \de v'$ is the overlap surface of two off-center circular 
rings with thickness $\de a$ and $\de b$, as shown in Fig. \ref{f.circ2}. 
It is shown easily by simple geometry that 
\beq \label{a.dvv2}
\de v\cap\de v'= \frac{2\de a \de b}{\sin (\theta +\phi)}, \quad (D=2),
\eeq
where $\theta$ and $\phi$ are defined in Fig. \ref{f.circ2} and expressed 
by Eqs (\ref{e.theta}) and (\ref{e.phi}). 
In the case $D=3$, $\de v \cap \de v'$ is the overlap volume of two spherical 
shells with internal radii $a$ and $b$, thickness $\de a$  and $\de b$, and
centers separated a distance $r$. This intersection volume is a torus with 
section area $\de a \de b/\sin(\theta+\phi)$ and perimeter $2\pi R$, where 
$\theta$ and $\phi$ are defined as before and $R=a\sin\theta=b\sin\phi$. 
Therefore, we can write
\beq \label{a.dvv3}
\de v\cap\de v'= \frac{2\pi a\sin\theta \de a \de b}{\sin (\theta +\phi)},
\quad (D=3).
\eeq
Equation (\ref{e.eta}) at $D=2$ and $D=3$ results from (\ref{eta_def}), 
(\ref{a.dvv2}) and (\ref{a.dvv3}), once the radii ($a$ and $b$) and thickness 
($\de a$ and $\de b$) are written in terms of $v=\alpha a^D$, $v'=\alpha b^D$,
$\de v=D\alpha a^{D-1}\de a$ and $\de v'=D\alpha b^{D-1}\de b$, with
$\alpha$ given by (\ref{e.alpha}).

\section{Partial probability densities at $D=1$} 
\label{a.D1}

For one-dimensional Poisson ensembles ($D=1$), Eqs. (\ref{gvv_1D}),
(\ref{ggv_nu1}), and (\ref{ggv_nu2}) yield
\beqn
g^{(i)}_v(\omega)&=&\frac 1n e^{-nv/2} \delta(\omega-v),\\
g^{(iia)}_v(\omega)&=&\frac 1n \left(1-e^{-nv/2}\right) \delta(\omega-v),\\
g^{(iib)}_v(\omega)&=&\frac 12 e^{-n\omega+nv/2} \Theta(\omega-v),\\
g^{(iii)}_v(\omega)&=&\frac 12 e^{-n(\omega-v)} \Theta(\omega-v),\\
g^{(iv)}_v(\omega)&=&\left[1- \frac 12 \left( e^{-n(\omega-v)}
+e^{-n\omega+nv/2}\right) \right] \Theta(\omega-v),\cr
&&
\eeqn
and
\beqn
g^{(i)}(\omega) &=& e^{-3n\omega/2},  \\
&&\cr
g^{(iia)}(\omega) &=& g^{(iib)}(\omega) = e^{-n\omega}-e^{-3n\omega/2},  \\
&&\cr
g^{(iii)}(\omega) &=& \frac {n\omega}2 e^{-n\omega}, \\
&&\cr
g^{(iv)}(\omega) &=& 1 -\frac {n\omega}2 e^{-n\omega}
                   -2 e^{-n\omega} + e^{-3n\omega/2}, 
\eeqn
Clearly, the sum of terms $g^{(\nu)}_v(\omega)$ and $g^{(\nu)}(\omega)$
over all pair classes ($\nu=$i, ii$a$, ii$b$, iii and iv) let us recover 
the exact expressions for $g_v(\omega)$ and $g(\omega)$, 
Eqs. (\ref{gv_poisson}) and (\ref{g_poisson}), respectively.
%
\section{Partial probability densities at $D=\infty$} 
\label{a.Dinf}

For Poisson ensembles at infinity dimensions, Eqs. (\ref{gvv_1D}),
(\ref{ggv_nu1}), and (\ref{ggv_nu2}) yield
\beqn 
\label{a.i1}
g^{(i)}_v(\omega)&=&\frac 1n e^{-nv} \delta(\omega-v), \\
\label{a.i2}
g^{(iia)}_v(\omega)&=&\frac 1n \left(1-e^{-nv}\right) \delta(\omega-v), \\
\label{a.i3}
g^{(iib)}_v(\omega)&=& e^{-n\omega} \Theta(\omega-v),\\
\label{a.i4}
g^{(iii)}_v(\omega)&=&0,\\
\label{a.i5}
g^{(iv)}_v(\omega)&=&\left( 1-e^{-n\omega} \right) \Theta(\omega-v),
\eeqn
and
\beqn 
\label{a.i6}
g^{(i)}(\omega)&=&e^{-2n\omega}, \\
\label{a.i7}
g^{(iia)}(\omega)&=& g^{(iib)}(\omega)= e^{-n\omega} -e^{-2n\omega},\\
\label{a.i8}
g^{(iii)}(\omega)&=&0,\\
\label{a.i9}
g^{(iv)}(\omega)&=&1-2 e^{-n\omega} +e^{-2n\omega}.
\eeqn
As in the case $D=1$ (Appendix \ref{a.D1}), Eqs. (\ref{gv_poisson}) and 
(\ref{g_poisson}) are recovered from (\ref{a.i1}-\ref{a.i5}) and 
(\ref{a.i6}-\ref{a.i9}), 
respectively.\\

%
\section{Attractive hard spheres in the TSP approach} 
\label{a.TSP}

The model of attractive hard spheres can be used to check the role played
by the CPDF in the evaluation of the potential energy contribution within 
the TSP approach.
We consider a pair interaction potential which consists of a hard core 
repulsion together with an attractive square well,
\beq \label{a.adh}
u(\omega)=\left\{
\begin{array}{rl}
 \infty,    & \quad \omega \le a, \\
 -\epsilon, & \quad a < \omega < a+\xi, \\
 0,         & \quad \omega \ge a+\xi , \\
\end{array}
\right.
\eeq
with $a$, $\xi$ and $\epsilon$ constants. For the present purpose, we may use
the expression of $g_v(\omega)$ given in \cite{Ro05} [eq. (41)] at the low 
density limit, i.e.,
\beq
g_v(\omega)=\frac 1n \delta(\omega-v-a^*)+\Theta(\omega-v-a^*),
\eeq
where $\Theta(x)$ is the Heaviside step function and $a^*=a/2$ is a reduction 
of the AV of hard particles in the limit of low density. 
Thus, with (\ref{a.adh}) into Eq. (\ref{e.phi1}) we obtain
\beq
\phi_v=\left\{
\begin{array}{ll}
 +\infty,    & \quad v \le b, \\
 -\epsilon\left[1+n(b+\xi-v) \right], & \quad b < v < b+\xi, \\
 \quad0,         & \quad \omega \ge b+\xi , \\
\end{array}
\right.
\eeq
where $b=a-a^*$.
With help of Eqs. (\ref{c.N}) and (\ref{c.V}) in the thermodynamic limit
($N,V\rightarrow \infty$ with $N/V=n$ constant) one finds
\beq
n_v= \frac{n\beta \gamma}{B} e^{\beta[\gamma(v-b)+\phi_v]}
\eeq
and
\beq \label{a.bg}
\beta\gamma= n\frac{B}{C},
\eeq
with
\beq \label{a.B}
B=e^{-\beta\gamma\xi}+\beta\gamma e^{\beta\epsilon(1+n\xi)}
\left[\frac{1-e^{-\beta(\gamma+n\epsilon)\xi}}{\beta(\gamma+n\epsilon)}\right]
\eeq
and
\beqn
C&=&[1+\beta\gamma(b+\xi)]e^{\beta\gamma\xi}+\left[\frac{\beta\gamma}
{\beta(\gamma+n\epsilon)}\right]^2e^{\beta\epsilon(1+n\xi)} \cr
&\times&
\{1+\beta(\gamma+n\epsilon)b-\left[1+\beta(\gamma+n\epsilon)(b+\xi)\right]
\cr
&\times&
e^{-\beta(\gamma+n\epsilon)\xi} \}
\eeqn
The contact value $g(\omega=a^+)$ of the PDF given in Eq. (\ref{a.gcont}) 
is directly obtained from Eq. (\ref{ggv}).


\end{document}